\documentclass[12pt]{article}
\usepackage{amssymb,amsmath,epsfig}
\usepackage{geometry}
\begin{document}

\title{\bf  Models of Anisotropic Self-Gravitating Source in Einstein-Gauss-Bonnet Gravity}
\author{G. Abbas \thanks{ghulamabbas@iub.edu.pk} and  M. Tahir \thanks{tmuhammad0064@gmail.com}
\\Department of Mathematics, The Islamia University \\of Bahawalpur, Bahawalpur-63100, Pakistan.}
\date{}
\maketitle
\begin{abstract}
In this paper, we have studied gravitational collapse and expansion of non-static anisotropic fluid in $5D$ Einstein Gauss-Bonnet gravity. For this purpose, the field equations have been modeled and evaluated for the given source and geometry. The two metric functions have been expressed in terms of parametric form of third metric function. We have examined the range of parameter $\beta$ (appearing in the form of metric functions) for which $\Theta$ the expansion scalar becomes positive/negative leads to expansion/collapse of the source. The trapped surface condition has been explored by using definition of Misner-Sharp mass and auxiliary solutions. The auxiliary solutions of the field equations involve a single function which generates two types of anisotropic solutions. Each solution can be represented in term of arbitrary function of time, this function has been chosen arbitrarily to fit the different astrophysical time profiles. The existing solutions forecast gravitational expansion and collapse depending on the choice of initial data. In this case, it has been investigated wall to wall collapse of spherical source. The dynamics of the spherical source has been observed graphically with the effects of Gauss-Bonnet coupling term $\alpha$ in the case of collapse and expansion. The energy conditions are satisfied for the specific values of parameters in the both solutions, this implies that the solutions are physically acceptable.
\end{abstract}
\section{Introduction}
  In the gravitational study of more than four dimensions the unification problem of gravity with electromagnetism and other basic connections are discussed by \cite{1,2}. The study on supergravity by Witten \cite{3}, has strongly sported the work on unification problem of gravity. The problem of gravitational unification study entirely based on the string theory \cite{4, 5}. The ten dimensional gravity that arises from string theory contains a quadratic term in its action \cite{7, 8}, in low energy limit. Zwiebach \cite{9}, described the ghost-free nontrivial gravitational interactions for greater than 4-dimensions in the study of $n$-dimensional action in \cite{7}.\\
In recent years, the higher order gravity included higher derivative curvature terms which is interesting and developing study. The most widely studied theory in the higher curvature gravities is known as Einstein Gauss-Bonnet (EGB) theory of gravity. The EGB theory of gravity is special case of the Lovelocks gravity. The Lagrangian of EGB gravity obtained from the first three terms of Lagrangian of the Lovelock theory. Pedro \cite{10}, described the $2^{nd}$ Euler density is the Gauss-Bonnet combination and is topological invariant in  four dimensions. He also pointed out that to make dynamical GB combination in four dimensional theory, couple it to dynamical scalar field. It has been studied in \cite{11,12}, the dynamical stability and adiabatic and anisotropic fluid collapse of stars in $5D$ EGB theory of gravity. Gross and Sloan \cite{13}, investigated that EGB theory of gravity occur in the low energy effectual action \cite{8} of super heterotic sting theory. The exact black hole (BH) solution in greater than or equal to five dimensions gravitational theories is studied by \cite{8}. Dadhich \cite{14}, examined in Newtonian theory that gravity is independent of spacetime dimensions with constant density of static sphere, and this result is valid for Einstein and higher order EGB theory of gravity. The conditions for universality of Schwarzschild interior solution describe sphere with uniform density for the dimensions greater than or equal to four. The authors \cite{15}-\cite{18} described that the existence of EGB term in the string theory lead to singularity free solutions in cosmology and hairy black holes. The solutions described in \cite{8} are generalization of $n$-dimensional spherically symmetric BH solution determined in \cite{19, 20}. In the literature \cite{21, 22, 23}, authors examined the other spherical symmetric BH solutions in GB theory of gravity. Cai \cite{24} discussed about the structure of topologically nontrivial black holes. The effects of GB term on the Vaidya solutions have been studied in \cite{25}-\cite{28}. Wheeler \cite{21}, discussed about the spherical symmetric BH solutions and their physical properties. The GB term has no effect on the existence of local naked singularity, but the strength of the curvature is effected.\\
In the composition of stars, the nuclear matter is enclosed inside the stars. The stars are gravitated and attracted continuously to the direction of its center because of gravitational interaction of its matter particles, this phenomenon is known as gravitational contraction of stars which leads to gravitational collapse. It is studied in \cite{29} that, during the gravitational collapse the space-time singularities are generated. When the massive stars collapse due to its own gravity, the end state of this collapse may be a neutron star, white dwarf, a BH or a naked singularity \cite{30}. The spherical symmetric collapse of perfect fluid is discussed in \cite{31, 32}. The dissipative and viscous fluid gravitational collapse in GR is discussed in literature \cite{33}-\cite{44}.\\
Zubair et al.\cite{b} studied a dynamical stability of cylindrical symmetric collapse of sphere filled with locally anisotropic fluid in $f(R, T)$ theory of gravity. A lot of literature is available about the gravitational collapse and BH in GB theory of gravity \cite{Ch1}-\cite{Ch8}. If the $f(R,T)$ theory of gravity obey the stress-energy tensor conservation then unknown $f(R,T)$ function can be obtained in the closed form \cite{c}.
Abbas and Riaz \cite{45}, determined the exact solution of non-static anisotropic gravitational fluid in $f(R,T)$ theory of gravity, that may leads to collapse and expansion of the star. Sharif and Aisha \cite{45a} studied the models for collapse and expansion of charged self gravitating objects in $f(R, T)$ theory of gravity.

Oppenheimer and Snyder \cite{46}, observed the gravitational contraction of in homogeneous spherically symmetric dust collapse, and according to this end state of the gravitational collapse is BH. Markovic and Shapiro \cite{50}, studied this work for positive cosmological constant, and Lake \cite{51} discuss this for negative as well as positive cosmological constant. Sharif and Abbas \cite{52} studied the gravitational perfect fluid charged collapse with cosmological constant in the Friedmann universe model with weak electromagnetic field. Sharif and Ahmad \cite{53}-\cite{56}, worked on the spherical symmetric gravitational collapse of perfect fluid withe positive cosmological constant. sharif and Abbas \cite{57}, discussed the 5-dimenssional symmetric spherical gravitational collapse with positive cosmological constant in the existence of an electromagnetic field. Abbas and Zubair \cite{g1} investigated the dynamical anisotropic gravitational collapse in EGB theory of gravity. A homogeneous spherical cloud collapse with zero rotation and disappearing internal pressure leads to a singularity covered by an event horizon \cite{g2}. Jhingan and Ghosh \cite{59} discussed the five or greater than five dimensional gravitational inhomogeneous dust collapse in EGB theory of gravity. They investigated the exact solution in closed form. Sunil et. al \cite{60}, investigated the exact solution to the field equations for five dimensional spherical symmetric and static distribution of the prefect fluid in EGB modified gravity . Abbas and Tahir \cite{61}, studied the exact solution of motion during gravitational collapse of prefect fluid in EGB theory of gravity. It should be observed in \cite{59} and \cite{61}-\cite{64} that coupling term $\alpha$ change the structure of the singularities. Glass \cite{65} generated the collapsing and expansion solutions of anisotropic fluid of Einstein field equations. In this paper, We extended the work of Glass \cite{65} to modeled the solutions for collapse and expansion of anisotropic fluid in the EGB theory of gravity. The paper has been arranged as: \\

In section $\textbf{2}$, we present interior matter distribution and field equations. We have discussed the generation of solutions for the gravitational collapse and expansion in section $\textbf{3}$. The last section present the summary for the results of this paper.

\section{Matter Distribution Inside the Star and the Field Equations}
We start with the $5D$ action given as:
\begin{eqnarray}\label{a1}
&&S= \int d^5 x \sqrt {-g}    \bigg[\frac{1}{2\kappa^2_{5}}\bigg(R+\alpha L_{GB}\bigg)\bigg]+S_{matter},
\end{eqnarray}
Where $\kappa_{5} \equiv \sqrt{8 \pi G_{5}} $ is gravitational constant and $R$ is a Ricci scalar in $5D$, and $\alpha$ is known as the coupling constant of the GB term. The GB Lagrangian is given below:
\begin{eqnarray}\label{a2}
 L_{GB} = R^2 -4R_{ab} R^{ab} +R_{abcd} R^{abcd}.
\end{eqnarray}
 This kind of action is discussed in the low energy limit of supersting theory \cite{13}. In this paper, we consider only the case with $\alpha > 0 $. The action (\ref{a1}) gives the following field equations:
\begin{eqnarray}\label{a3}
  G_{ab} + \alpha H_{ab} = T_{ab},
\end{eqnarray}
where
\begin{eqnarray}\label{a4}
 &&G_{ab}=R_{ab} - \frac{1}{2} g_{ab}R
\end{eqnarray}
is the Einstein tensor and
\begin{eqnarray}\label{a5}
 &&H_{ab}=2\bigg(RR_{ab} -2 R_{a \alpha} R^{\alpha}_{b} -2R^{\alpha \beta} R_{a \alpha b \beta}+R^{\alpha \beta \gamma}_{a} R_{b \alpha \beta \gamma}\bigg) -\frac{1}{2} g_{ab} L_{GB}
\end{eqnarray}
is the Lanczos tensor. We want to find the solution for collapse and expansion of a spherical anisotropic fluid in 5D-EGB gravity.
\begin{eqnarray}\label{a6}
 &&ds^2 = -A\bigg(t,r\bigg)^2dt^2 +B\bigg(t,r\bigg)^2 dr^2 + R\bigg(t,r\bigg)^2 d\Omega^2_{3},
\end{eqnarray}
where $d\Omega^2_{3} = (d\theta^2 +\sin^2\theta (d\phi^2+\sin^2\phi d\psi^2))$, is a metric on three-sphere, and $R = R(t, r)\geq0$, $A=A(t,r)$ and $B=B(t,r)$.The energy-momentum tensor for anisotropic fluid is
\begin{eqnarray}\label{a7}
&&T_{ab}= \bigg(\mu+p_{\bot}\bigg)V_{a}V_{b}+p_{\bot}g_{ab}+\bigg(p_{r}-p_{\bot}\bigg)\chi_{a}\chi_{b} ,
\end{eqnarray}
where $\mu$, $p_{r}$, $p_{\bot}$, $\chi_{a}$ and $V_{a}$ are the energy density, radial pressure, tangential pressure, unit four vector along the radial direction and four velocity of the fluid respectively. For the metric (\ref{a6}) $V_{a}$ and $\chi_{a}$ are given by \cite{45} are:
\begin{eqnarray}\label{a9}
 &&V^{a}=A^{-1}\delta^{a}_{0}  ,~~~~\chi^{a}=B^{-1}\delta^{a}_{1}
\end{eqnarray}
which satisfy
\begin{eqnarray}\label{b1}
 &&V^{a}V_{a}=-1   ,~~~   \chi^{a}\chi_{a}=1     , ~~~   \chi^{a}V_{a}=0.
\end{eqnarray}
The expansion scalar is
\begin{eqnarray}\label{b2}
 &&\Theta=\frac{1}{A}\bigg(\frac{\dot{B}}{B}+\frac{3\dot{R}}{R}\bigg).
\end{eqnarray}
The dimensionless measure of anisotropy is defined as
\begin{eqnarray}\label{b3}
 &&\triangle a=\frac{p_{r}-p_{\bot}}{p_{r}}.
\end{eqnarray}
The Eq.(\ref{a3}) for the metric Eq.(\ref{a6}) with the help of Eq.(\ref{a7}), are given bellow:
\begin{eqnarray}\label{b4}
 &&\mu=\nonumber\frac{12\alpha}{A^4B^5R^3}\bigg[\dot{R}\bigg(\dot{R}\bigg(A'^2R-B^2\dot{B}\dot{R}+A^2\bigg(B R''-B'R'\bigg)\bigg)B^2
 +A^2B^2\dot{B}\bigg(R'^2-B^2\bigg)\bigg)
\\&&\nonumber +2AA'B^2R\dot{R}\bigg(\dot{B}R'-B\dot{R}'\bigg)
 +A^4\bigg(B^2-R'^2\bigg)\bigg(B R''-B'R'\bigg)\bigg]-\frac{3}{A^2B^3R^2}\bigg[B^3\bigg(A^2+\dot{R}^2\bigg)\\&&
 +A^2R'\bigg(B'R-B R'\bigg)+B R\bigg(B\dot{B}\dot{R}-A^2R''\bigg)\bigg],
\end{eqnarray}
\begin{eqnarray}\label{b5}
&&p_{r}=\nonumber\frac{12\alpha}{A^5B^4R^3}\bigg[B^2\dot{R}^2\bigg(\dot{A}\dot{R}B^2-A\bigg(2A'^2R+B^2\ddot{R}\bigg)\bigg)
+A^2\bigg(\bigg(A^2A'R'\\&&\nonumber
+B^2\bigg(\dot{A}\dot{R}-A\ddot{R}\bigg)\bigg)
\bigg(R'^2-B^2\bigg)+B^2A'R'\dot{R}^2\bigg)\bigg]
+\frac{3}{A^3B^2R^2}\bigg[A^2\bigg(\bigg(A'R+AR'\bigg)R'\\&&
-AB^2\bigg)+B^2\bigg(\bigg(\dot{A}\dot{R}-A\ddot{R}\bigg)R-A\dot{R}^2\bigg)\bigg],
\end{eqnarray}
\begin{eqnarray}\label{b6}
&&p_{\perp}=\nonumber\frac{-4\alpha}{A^5B^5R^2}\bigg[B^3\dot{B}\bigg(3B\dot{A}\dot{R}^2+A\bigg(A'^2\dot{R}
-B\bigg(\ddot{B}\dot{R}+2\dot{B}\ddot{R}\bigg)\bigg)\bigg)
+\bigg(AB\bigg)^2\bigg(\dot{R}\bigg(BA''\dot{R}\\&&\nonumber
+A'\bigg(4\dot{B}R'-2B\dot{R}'-B'\dot{R}\bigg)\bigg)
+\dot{A}\bigg(\bigg(B^2-R'^2\bigg)\dot{B}
+2\dot{R}\bigg(B'R'-B R''\bigg)\bigg)\bigg)\\&&\nonumber
+A^3B\bigg(\dot{B}^2R'^2-B^3\ddot{B}+BR'\bigg(\ddot{B}R'-2B'\ddot{R}-2\dot{B}\dot{R}'\bigg)
+B^2\bigg(\dot{R}'^2+2\ddot{R}R''\bigg)\bigg)\\&&\nonumber
+A^4\bigg(A''B\bigg(B^2-R'^2\bigg)-A'\bigg(B\bigg(B B'+2R'R''\bigg)-3B'R'^2\bigg)\bigg)\bigg]\\&&\nonumber
-\frac{1}{\bigg(A B\bigg)^3R^2}\bigg[R^2\bigg(\bigg(A\ddot{B}-\dot{A}\dot{B}\bigg)B^2
+A^2\bigg(A'B'-A''B\bigg)\bigg)
+2R\bigg(AB\bigg(\bigg(\dot{B}\dot{R}+B\ddot{R}\bigg)B\\&&
-A\bigg(A'R'+AR''\bigg)\bigg)
+A^3B'R'-B^3\dot{A}\dot{R}\bigg)
+A B\bigg(B^2\bigg(A^2+\dot{R}^2\bigg)-A^2R'^2\bigg)\bigg],
\end{eqnarray}
\begin{eqnarray}\label{b7}
&&0=\nonumber\frac{-12\alpha}{A^4B^4R^3}\bigg(A'B\dot{R}-AB\dot{R}'+A\dot{B}R'\bigg)\bigg(2A^2\bigg(A'B'-A''B\bigg)R^2
-B^2\bigg(AB\dot{R}^2+2\dot{A}\dot{B}R^2\bigg)\\&&
+A^3B\bigg(R'^2-B^2\bigg)\bigg)
+\frac{3}{A B R}\bigg(A'B\dot{R}-AB\dot{R}'+A\dot{B}R'\bigg)
\end{eqnarray}
 where the prime and dot denotes the partial derivative with respect to r and t respectively.
The auxiliary solution of the Eq.(\ref{b7}) is
\begin{eqnarray}\label{b8}
&&A=\frac{\dot{R}}{R^{\beta}},       B=R^{\beta}.
\end{eqnarray}
By Using the Auxiliary solution (\ref{b8}) in Eq.(\ref{b2}), the expansion scalar becomes
 \begin{eqnarray}\label{b9}
&&\Theta=R^{\beta-1}\bigg(3+\beta\bigg).
\end{eqnarray}
For $\beta>-3$ and $\beta<-3$ , we obtained expansion and collapse regions.
The matter components from Eq.(\ref{b4}), Eq.(\ref{b5}) and Eq.(\ref{b6}) with the help of Eq.(\ref{b8}) are
\begin{eqnarray}\label{c1}\nonumber
&&\mu= \frac{12\alpha}{R^{4(1-\beta)}}\bigg[R^{2\beta}\bigg(\bigg(\bigg(\bigg(\frac{2R\dot{R}'}{\dot{R}}-\beta R'\bigg)R'-R^{2\beta}-1\bigg)R^{4\beta}-\beta R'^{2}\bigg)+R'^{4}\bigg)\beta
+R^{2\beta}\bigg(1+R^{2\beta}\bigg)\\&&
-R'^{2}\bigg)RR''\bigg)-\frac{R^{4\beta}R^{2}\dot{R}'^{2}}{\dot{R^{2}}}\bigg]
-\frac{3}{R^{2(1+\beta)}}\bigg[\bigg(\beta -1\bigg)R'^{2}-RR''+R^{2\beta}\bigg(1+R^{2\beta}\bigg(1+\beta\bigg)\bigg],
\end{eqnarray}
\begin{eqnarray}\label{c2}
&&p_{r}=\nonumber\frac{12\alpha}{R^{4(1-\beta)}}\bigg[R^{4\beta}\bigg(R^{2\beta}\bigg(1-R^{2\beta}\bigg)
+R'\bigg(\frac{4R\dot{R}'}{\dot{R}}+R'\bigg(\frac{1}
{R^{2\beta}}+\frac{R'^2}{R^{4\beta}}-2\bigg)\bigg)\bigg)\beta\\&&\nonumber
+\frac{RR'\dot{R}'}{\dot{R}}\bigg(\bigg(1-R^{2\beta}\bigg)R^{2\beta}-R'^{2}\bigg)
-2\bigg(\frac{R^{2\beta}R\dot{R}'}{\dot{R}}\bigg)^2]
+\frac{3}{R^2}\bigg[\frac{1}{R^{2\beta}}\bigg(R'^2-R\ddot{R}-\dot{R}^2\bigg)\\&&
+\frac{1}{R^{4\beta}}\bigg(R\dot{R}\ddot{R}+\acute{R}\acute{\dot{R}}
-\beta\dot{R}\bigg(\dot{R}^2+\acute{R}^2\bigg)\bigg)-1\bigg],
 \end{eqnarray}
\begin{eqnarray}\label{c3}\nonumber
&&p_{\bot}=\nonumber\frac{4\alpha}{R^{4(1+\beta)}}\bigg[R^{4\beta}\bigg(\beta\bigg(\bigg(4\beta-1\bigg)R^{4\beta}
+R^{2\beta}\bigg(2\beta-1\bigg)\bigg)-\beta R'' +R'^{2}\bigg(\beta^{2}-1\bigg)\bigg)\\&&\nonumber
+R^{2\beta}\bigg(R R''\beta-R'^{2}\bigg(1+\beta+\beta^{2}\bigg)\bigg)+R'^{2}(R'^{2}\bigg(1+\beta+3\beta^{2}\bigg)-3\beta R R''\bigg)\\&&\nonumber
+\frac{R^{2\beta}R}{\dot{R}}\bigg(\bigg(\dot{R}' R'\beta-R\dot{R}''\bigg)\bigg(R^{2\beta}+1\bigg)\bigg)+\frac{RR'}{\dot{R}}\bigg(R'\bigg(R\dot{R}''-5\beta \dot{R}R'\bigg)+2R\dot{R}'R''\bigg)\bigg]\\&&\nonumber
-\frac{1}{R^{2(1+\beta)}}\bigg[R^{2\beta}-R'^{2}\bigg(\beta^{2}-3\beta+2\bigg)+R^{4\beta}\bigg(2\beta^{2}
+3\beta+1\bigg)+RR''\bigg(\beta-2\bigg)\\&&
+\frac{R}{\dot{R}}\bigg(R'\dot{R}'\bigg(3\beta-2\bigg)-R\dot{R}''\bigg)\bigg].
\end{eqnarray}
The Misner-Sharp mass function m(t,r) is given below
 \begin{eqnarray}\label{c4}
m\bigg(t,r\bigg)=\frac{\bigg(n-2\bigg)}{2k^{2}_{n}}V^{k}_{n-2}\bigg(R^{n-3}\bigg(k-g^{ab}R_{, a}R_{, b}\bigg)+\bigg(n-3\bigg)\bigg(n-4\bigg)\alpha\bigg(k-g^{ab}R_{,a}R_{,b}\bigg)^{2}\bigg)
\end{eqnarray}
where comma represents partial differentiation and the surface of $(n-2)$ dimensional unit space is represented by $V^{k}_{n-2}$ . By using $V^{1}_{n-2}=\frac{2\pi(n-1)/2}{\Gamma((n-1)^2)}$, $k=1$ with $n=5$ and Eq.(\ref{b8}) in Eq.(\ref{c4}) we get
\begin{eqnarray}\label{c5}
&&m\bigg(r,t\bigg)=\frac{3}{2}\bigg(R^2\bigg(1-\frac{R'^2}{R^{2\beta}}+R^{2\beta}\bigg)+2\alpha\bigg(1-\frac{R'^2}{R^{2\beta}}
+R^{2\beta}\bigg)^2\bigg).
\end{eqnarray}
The specific values of $\beta$ and $R(r,t)$ form an anisotropic configuration.
When $R'=R^{2\beta}$ , there exists trapped surfaces at $R=\pm\sqrt{\frac{2}{3}m-2\alpha}$ , provided $m\geq3\alpha$. Therefore, in this case $R'=R^{2\beta}$ is trapped surface condition.
\section{Generating Solutions}
For the different values of $\beta$, expansion scalar $\Theta<0$ for collapse and $\Theta>0$ for expansion solutions which are discussed as follows:
\section{Collapse with $\beta=-\frac{7}{2}$}
The expansion scalar must be negative in case of the collapse, so for $\beta<-3$, from Eq.(\ref{b9}), we get $\Theta<0$. By assuming $\beta=-\frac{7}{2}$ and the trapped condition $R'=R^{2\beta}$, and then integrate, we obtained,
\begin{eqnarray}\label{c6}
&&R_{trap}=\bigg(8r+z_{1}(t)\bigg)^{\frac{1}{8}}.
\end{eqnarray}
It is to noted that Eq.(\ref{c6}) is only valid for the trapped surface, and it can not be used every where. In order to discuss the solutions out side the trapped surface, we follow Glass \cite{65} and take the value of $R(r,t)$ as posivite scalar ($k>1$) multiple of $R_{trap}$, such that $R(r,t)>R_{trap}$. Hence the convenient form of the areal radius $R(r,t)$ for the solution out the trapped surface is given by
\begin{eqnarray}\label{t1}
&&R=k\bigg(8r+z_{1}(t)\bigg)^{\frac{1}{8}},~~~~~and~~~ z_{1}(1)=1+t^2.
\end{eqnarray}

Thus Eqs.(\ref{c1})-(\ref{c3}) with Eq.(\ref{t1}) and $\beta=-\frac{7}{2}$ yield the following set of equations
\begin{eqnarray}\label{t2}
&&\mu=\frac{-21\alpha}{k^{18}\bigg(8r+z_{1}\bigg)^{\frac{9}{4}}}\bigg[-2+k^{7}\bigg(\bigg(11-2k^{16}\bigg)k^9
-2\bigg(8r+z_{1}\bigg)^{\frac{7}{8}}\bigg(1-k^{25}\bigg)\bigg)\bigg]\nonumber\\&&
+\frac{3}{2k^9\bigg(8r+z_{1}\bigg)^{\frac{9}{8}}}\bigg[5\bigg(k^{16}-1\bigg)+2k^{7}\bigg(8r+z_{1}\bigg)^{\frac{7}{8}}\bigg],
\end{eqnarray}
\begin{eqnarray}\label{t3}
&&p_{r}=\frac{-42\alpha}{k^{18}\bigg(8r+z_{1}\bigg)^{\frac{9}{4}}}\bigg[-1+k^{7}\bigg(\bigg(9-k^{16}\bigg)k^9
-\bigg(8r+z_{1}\bigg)^{\frac{7}{8}}\bigg(1-k^{25}\bigg)\bigg)\bigg]\nonumber\\&&
-\frac{3}{1024k^2\bigg(8r+z_{1}\bigg)^{\frac{5}{4}}}\bigg[k^9\bigg(896k^7\dot{z_{1}}
+\bigg(8r+z_{1}\bigg)^{\frac{1}{8}}\bigg(1024+96\dot{z_{1}}^2-k^8\dot{z_{1}}\nonumber\\&&
\bigg(128+7\dot{z_{1}}^2\bigg)\bigg)\bigg)+\bigg(128r+16z_{1}\bigg)\bigg(\bigg(8r+z_{1}\bigg)^{\frac{1}{8}}\bigg(k^8\dot{z_{1}}
-8\bigg)k^9\ddot{z_{1}}
-64\bigg)\bigg],
\end{eqnarray}
\begin{eqnarray}\label{t4}
&&p_{\bot}=\frac{\alpha}{k^{18}\bigg(8r+z_{1}\bigg)^{\frac{9}{4}}}\bigg[210-k^{7}\bigg(\bigg(375-165k^{16}\bigg)k^9
-\bigg(8r+z_{1}\bigg)^{\frac{7}{8}}\bigg(112-67k^{25}\bigg)\bigg)\bigg]\nonumber\\&&
+\frac{1}{4k^9\bigg(8r+z_{1}\bigg)^{\frac{9}{8}}}\bigg[60-k^7\bigg(315k^9-4\bigg(8r+z_{1}\bigg)^{\frac{7}{8}}\bigg)\bigg].
\end{eqnarray}
In this case, the Eq.(\ref{c5}) is reduced to the following form
\begin{eqnarray}\label{d4}
&&m\bigg(r,t\bigg)=\frac{3}{2}\bigg[k^2\bigg(8r+z_{1}\bigg)^{\frac{1}{4}}-\frac{k^{11}}{\bigg(8r+z_{1}\bigg)^{\frac{5}{8}}}
+\frac{\bigg(8r+z_{1}\bigg)^{\frac{1}{8}}}{k^5}
+2\alpha\bigg(1-\frac{k^9}{\bigg(8r+z_{1}\bigg)^{\frac{7}{8}}}\nonumber\\&&
+\frac{1}{k^7\bigg(8r+z_{1}\bigg)^{\frac{7}{8}}}\bigg)\bigg].
\end{eqnarray}
The Eq.(\ref{b3}), with the help of Eq.(\ref{t3}) and Eq.(\ref{t4}) is
\begin{eqnarray}\label{d5}
&&\Delta a=1-\frac{C_{1}+C_{2}}{C_{3}+C_{4}}
\end{eqnarray}
where\\
$C_{1}=\frac{45\alpha}{\bigg(8r+z_{1}\bigg)^{\frac{11}{8}}}$,\\
$C_{2}=\frac{1}{\bigg(8r+z_{1}\bigg)^{\frac{1}{8}}}\bigg[\frac{45}{4\bigg(8r+z_{1}\bigg)}+1\bigg]$,\\
$C_{3}=-294\alpha\bigg(8r+z_{1}\bigg)^{-\frac{9}{4}}$\\
and\\
$C_{4}=-\frac{3}{1024}\bigg[\frac{896\dot{z_{1}}}{\bigg(8r+z_{1}\bigg)^{\frac{5}{4}}}
+\frac{1}{\bigg(8r+z_{1}\bigg)^{\frac{9}{8}}}\bigg(1024-128\dot{z_{1}}
+96\dot{z_{1}}^2-7\dot{z_{1}}^3\bigg)\\+\frac{\bigg(128r+16z_{1}\bigg)}{\bigg(8r
+z_{1}\bigg)^{\frac{5}{4}}}\bigg(\bigg(8r+z_{1})^{\frac{1}{8}}
\bigg(\dot{z_{1}}-8\bigg)\ddot{z_{1}}-64\bigg)\bigg].$\\
 \begin{figure}
\begin{center}
\includegraphics[width=70mm]{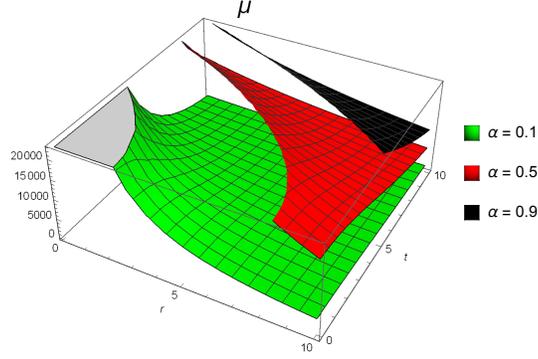}
\caption{Plot of density along $r$ and $t$ for $k=2.5$ and the different values of $\alpha$.}
\end{center}
\end{figure}
\begin{figure}
\begin{center}
\includegraphics[width=70mm]{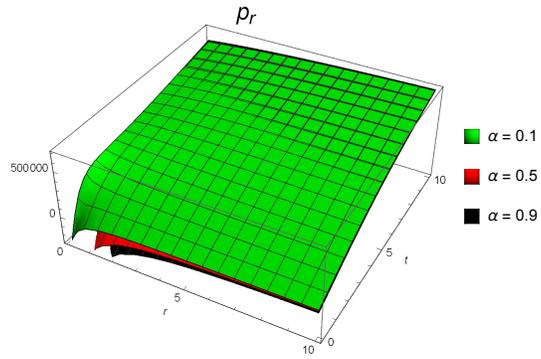}
\caption{Behavior of Radial pressure along $r$ and $t$ for $k=2.5$ and the different values of $\alpha$.}
\end{center}
\end{figure}
\begin{figure}
\begin{center}
\includegraphics[width=70mm]{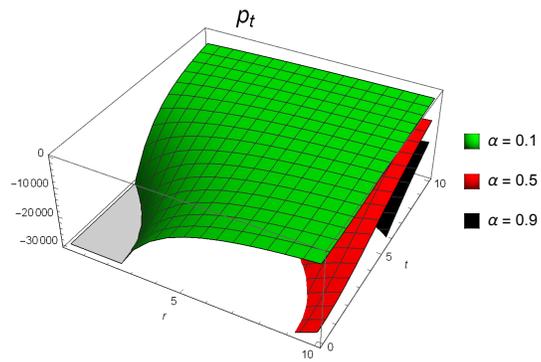}
\caption{Tangential pressure along $r$ and $t$ for $k=2.5$ and the different values of $\alpha$.}
\end{center}
\end{figure}
\begin{figure}
\begin{center}
\includegraphics[width=70mm]{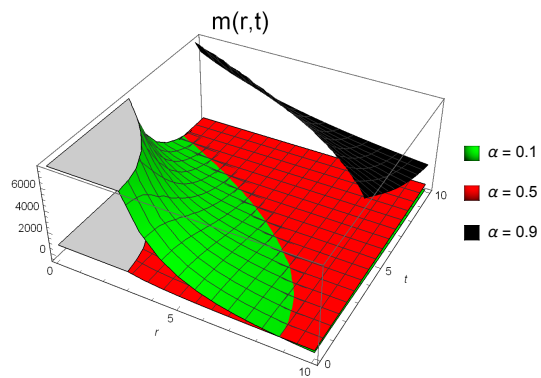}
\caption{Mass plot along $r$ and $t$ for $k=2.5$ and the different values of $\alpha$.}
\end{center}
\end{figure}
\begin{figure}
\begin{center}
\includegraphics[width=70mm]{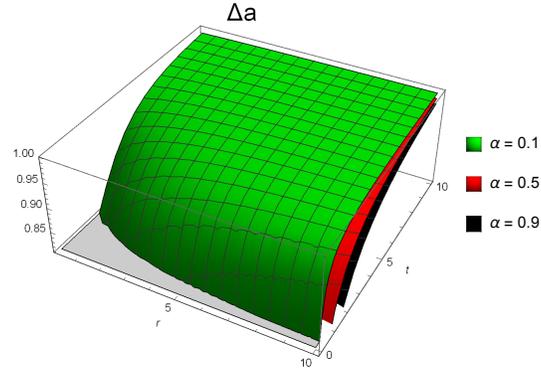}
\caption{Plot of Anisotropy parameter along $r$ and $t$ for $k=2.5$ and the different values of $\alpha$.}
\end{center}
\end{figure}
\begin{figure}
\begin{center}
\includegraphics[width=60mm]{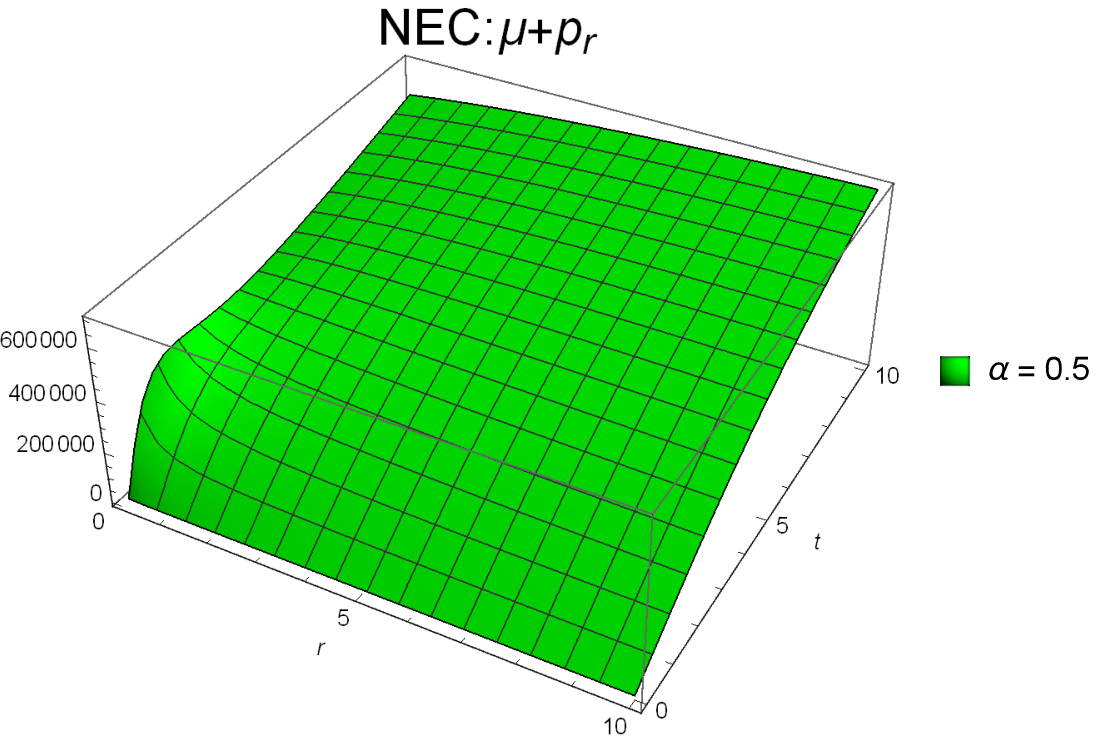}
\includegraphics[width=60mm]{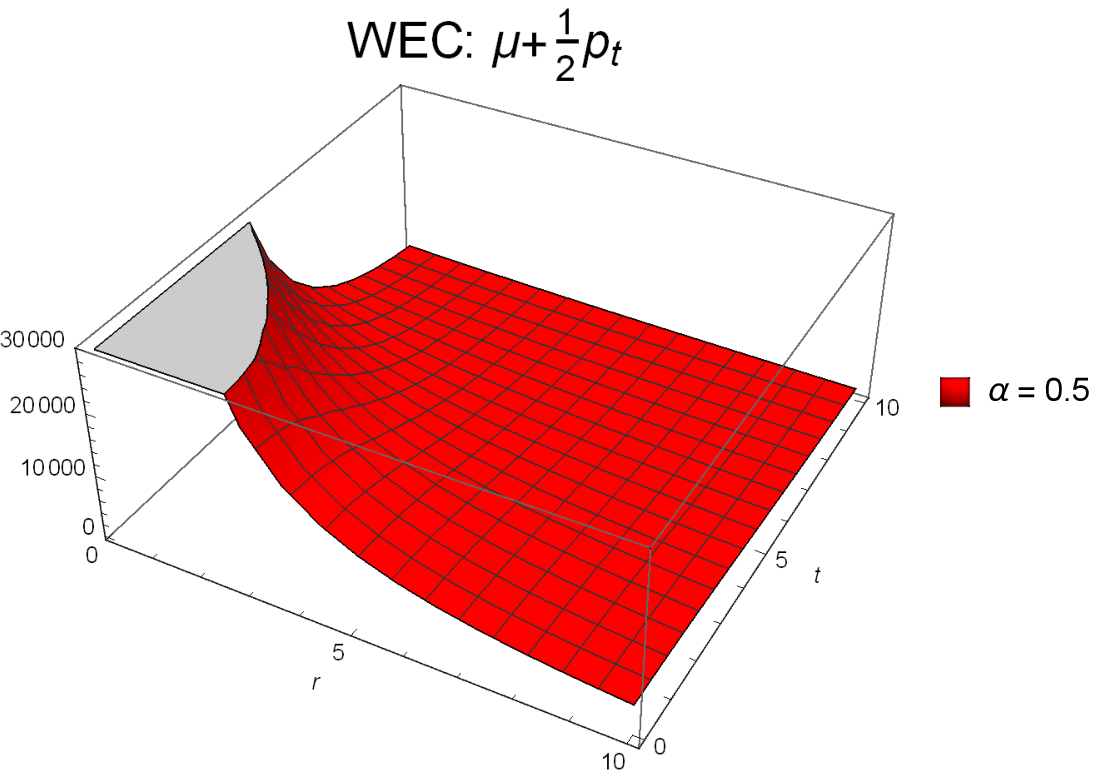}
\includegraphics[width=60mm]{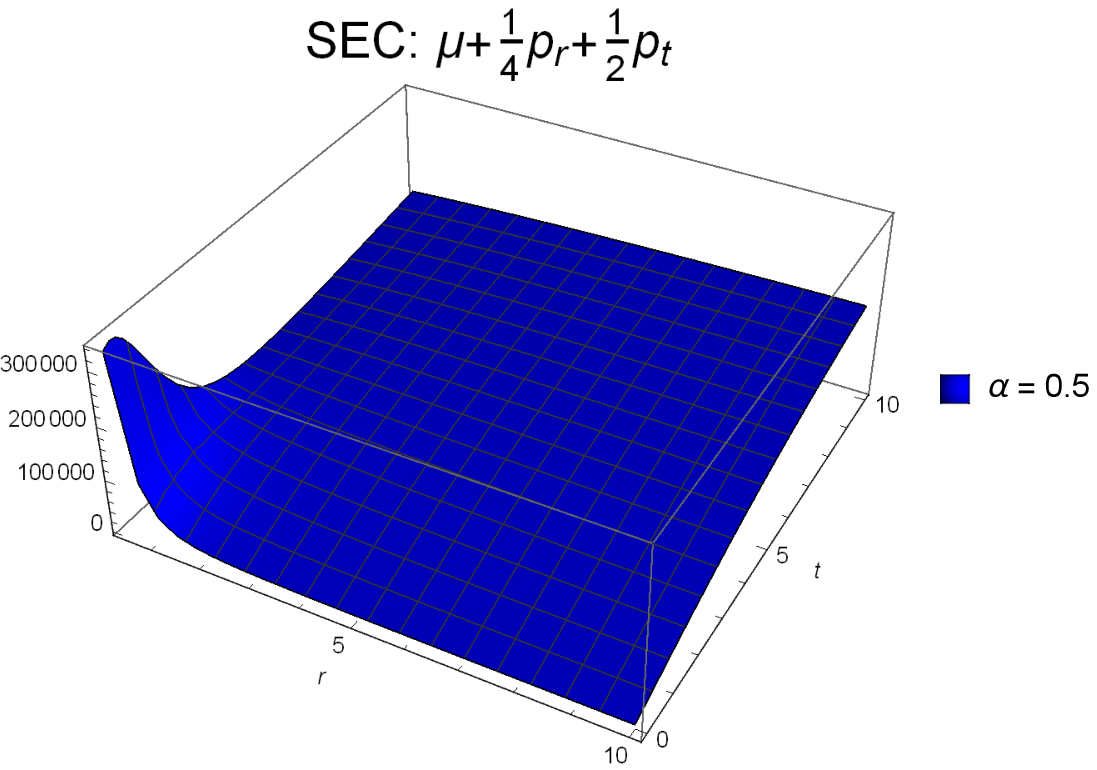}
\caption{Plot of energy conditions along $r$ and $t$ for $k=2.5$ .}
\end{center}
\end{figure}

The energy conditions for the curvature matter coupled gravity are\\

(i) Null energy condition:        $ \mu+p_{r}\geq0,~~~~\mu+\frac{1}{2}p_{t}\geq0$\\

(ii) Weak energy condition:       $\mu\geq0,~~~\mu+p_{r}\geq0,~~~~\mu+\frac{1}{2}p_{t}\geq0$\\

(iii) Strong energy condition:    $\mu+\frac{1}{4}p_{r}+\frac{1}{2}p_{t}\geq0$\\

In this case, we analyzed our results for time profile $z_{1}=1+t^2$, $k=2.5$ and the different values of $\alpha$.
For $\beta=-\frac{7}{2}$, we obtained $\Theta<0$, the energy density is decreasing with respect to radius r and time t, and remains positive for different values of coupling constant $\alpha$ as shown in \textbf{Fig(1)}. As the density is decreasing, so spherical object goes to collapse outward the point. It is observed in \textbf{Fig(2)}, that the radial pressure is increasing outward the center, and also noted in \textbf{Fig(3)} that, the tangential pressure is increasing with respect to radius r and time t at various values of $\alpha$. Due to this increase of radial and tangential pressure on the surface, the sphere loss the equilibrium state, which may cause the collapse of the sphere outward the center. The \textbf{Fig(4)} shows that, the mass is decreasing function of r and t during the collapse at different values of $\alpha$. The anisotropy directed outward when $p_{t}<p_{r}$, this implies that $\Delta a>0$ and directed inward when $p_{r}<p_{t}$, this implies that $\Delta a<0$. In this case $\Delta a>0$ for increasing r and t at various values of $\alpha$ as shown in \textbf{Fig(5)}. This represents that the anisotropy force allows the formation of more massive object and have attractive force for $\Delta a<0$ near the center. In the present case the GB coupling term $\alpha$ effect the anisotropy and the homogeneity of the collapsing sphere. All the energy conditions are plotted in \textbf{Fig(6)}, these plots represent that all the energy conditions are satisfied for considered parameters in collapse solutions.

\section{Expansion with $\beta=-\frac{5}{2}$}
In this case, the expansion scalar $\Theta$ must be positive. When $\beta>-3$, From Eq.(\ref{b9}) the expansion scalar $\Theta$ is positive. We assume that
\begin{eqnarray}\label{d6}
&&R=\bigg(r^2+r_{0}^{2}\bigg)^{-1}+z_{2}(t),
\end{eqnarray}
where $r_{0}$ is constant and we take $z_{2}=z_{2}(t)=1+t^2$.\\
For $\beta=-\frac{5}{2}$, Eq.(\ref{c1}), Eq.(\ref{c2}) and Eq.(\ref{c3}) take the form
\begin{eqnarray}\label{d7}
&&\mu=\alpha\bigg[\frac{30}{R^9}\bigg(\frac{1}{R^5}+1\bigg)+6R\bigg(5R'^2+2RR''\bigg)\bigg(1-R^5R'^2\bigg)
+\frac{1}{R^3}\bigg(12R''-75\frac{R'^2}{R}\bigg)\nonumber\\&&
-\frac{12\dot{R'}}{R^2\dot{R}}\bigg(\frac{1}{\dot{R}}+\frac{5R'}{R}\bigg)\bigg]
-\frac{3}{2}\bigg[\frac{1}{R^2}\bigg(\frac{3}{R^5}-2\bigg)+R^3\bigg(7R'^2+2RR''\bigg)\bigg],
\end{eqnarray}
\begin{eqnarray}\label{d8}
&&p_{r}=-6\alpha\bigg[\frac{2\dot{R'}}{R^2\dot{R}}\bigg(\frac{2\dot{R'}}{\dot{R}}+\frac{9R'}{R}\bigg)
+RR'\bigg(\frac{2R\dot{R'}}{\dot{R}}
+5\dot{R}^3R'\bigg)\bigg(R^5R'^2-1\bigg)\nonumber\\&&
+\frac{5\dot{R}^3}{R^{14}}\bigg(R^5\bigg(5R^5R'^2
-1\bigg)-1\bigg)\bigg]+\frac{3}{4}\bigg[2R^3R'\bigg(\frac{2R\dot{R'}}{\dot{R}}+7R'\bigg)\nonumber\\&&
-\frac{\ddot{R}}{R^6\dot{R}^2}\bigg(\frac{5}{R^{\frac{7}{2}}}+2\bigg)
-\frac{1}{R^2}\bigg(\frac{25}{R^{\frac{17}{2}}}+\frac{4}{R^5}+4\bigg)\bigg],
\end{eqnarray}
\begin{eqnarray}\label{d9}
&&p_{\bot}=\alpha\bigg[\frac{1}{R^4}\bigg(\frac{110}{R^{10}}+\frac{60}{R^5}+21R'^2\bigg)+RR'^2\bigg(69R^5R'^2-19\bigg)
+10R''\bigg(\frac{1}{R^3}
+R^2\bigg(3R^5R'^2-1\bigg)\bigg)\nonumber\\&&
+\frac{2}{R^2\dot{R}}\bigg(\dot{R}''\bigg(R^5\bigg(R^5R'^2-1\bigg)-1\bigg)
+\frac{R'\dot{R}'}{R}\bigg(R^5\bigg(4R^6R''
+25R^5R'^2-15\bigg)-5\bigg)\bigg)\bigg)\bigg]\nonumber\\&&
+\frac{1}{R^2}\bigg(\frac{6}{R^5}
+1\bigg)-\frac{9}{2}R^3\bigg(RR''+\frac{7}{2}R'^2\bigg)-\frac{R^4}{\dot{R}}\bigg(R\dot{R}''-\frac{19}{2}R'\dot{R}'\bigg).
\end{eqnarray}
Assuming that $F\bigg(r,t\bigg)=1+z_{2}(t)\bigg(r^2+r_{0}^{2}\bigg)$ and $R=\frac{F}{\bigg(r^2+r_{0}^{2}\bigg)}$, the Eq.(\ref{d7}), Eq.(\ref{d8}) and Eq.(\ref{d9}) in this case are,

\begin{eqnarray}\label{e1}
&&\mu=\alpha\bigg[10\bigg(\frac{r^2+r_{0}^2}{F}\bigg)^9\bigg(1+\bigg(\frac{r^2+r_{0}^2}{F}\bigg)^5\bigg)
+\frac{8F^2}{\bigg(r^2+r_{0}^2\bigg)^5}\bigg(\frac{4r^2F^5}
{\bigg(r^2+r_{0}^2\bigg)^9}-1\bigg)\bigg(F\bigg(5r^2+r_{0}^2\bigg)\nonumber\\&&
-z_{2}\bigg(r^2+r_{0}^2\bigg)\bigg(r_{0}^2-3r^2\bigg)\bigg)+4r^2\bigg(\frac{20F}{\bigg(r^2+r_{0}^2\bigg)^5}
-\frac{25}{F^4}-\frac{40r^2F^6}{\bigg(r^2+r_{0}^2\bigg)^{14}}\bigg)
-\frac{8\bigg(5r^2+r_{0}^2\bigg)}{F^2}\nonumber\\&&
+\frac{8z_{2}}{F^3}\bigg(r^2+r_{0}^2\bigg)\bigg(r_{0}^2-3r^2\bigg)\bigg]
-\frac{3}{2}\bigg[\bigg(r^2+r_{0}^2\bigg)^2\bigg(\frac{3\bigg(r^2+r_{0}^2)^5}{F^7}-\frac{2}{F^2}\bigg)\nonumber\\&&
+\frac{4F^3}{\bigg(r^2+r_{0}^2\bigg)^6}\bigg(\bigg(r_{0}^2-3r^2\bigg)z_{2}
-\frac{F^2\bigg(5r^2+r_{0}^2\bigg)}{\bigg(r^2+r_{0}^2\bigg)}+\frac{7r^2}{\bigg(r^2+r_{0}^2\bigg)}\bigg)\bigg],
\end{eqnarray}
\begin{eqnarray}\label{e2}
&&p_{r}=6\alpha\bigg[5\bigg(\frac{r^2+r_{0}^2}{F}\bigg)^9\bigg(1+\bigg(\frac{r^2+r_{0}^2}{F}\bigg)^5\bigg)
-4r^2\bigg(\frac{25}{F^4}-\frac{5F}{\bigg(r^2+r_{0}^2\bigg)^5}
\bigg(1-\frac{4r^2F^5}{\bigg(r^2+r_{0}^2\bigg)^9}\bigg)\bigg]\nonumber\\&&
+\frac{3}{4}\bigg[-2\frac{\ddot{z_{2}}}{\dot{z_{2}}}\bigg(\frac{r^2+r_{0}^2}{F}\bigg)^{\frac{19}{2}}\bigg(5
+2\bigg(\frac{F}{r^2+r_{0}^2}\bigg)^{\frac{7}{2}}\bigg)
+\frac{56r^2F^3}{\bigg(r^2+r_{0}^2\bigg)^7}-25\bigg(\frac{r^2+r_{0}^2}{F}\bigg)^\frac{21}{2}\nonumber\\&&
-4\bigg(\frac{r^2+r_{0}^2}{F}\bigg)^7-4\bigg(\frac{r^2+r_{0}^2}{F}\bigg)^2\bigg],
\end{eqnarray}
\begin{eqnarray}\label{e3}
&&p_{\bot}=\alpha\bigg[60\bigg(\frac{r^2+r_{0}^2}{F}\bigg)^9+110\frac{\bigg(r^2+r_{0}^2\bigg)^8}{F^{14}}
+\bigg(\frac{10F^2}{\bigg(r^2+r_{0}^2\bigg)^5}-\frac{10}{F^3}
-\frac{60r^2F^7}{\bigg(r^2+r_{0}^2\bigg)^{14}}\bigg)\bigg(2F\bigg(5r^2+r_{0}^2\bigg)\nonumber\\&&
-2z_{2}\bigg(r^2+r_{0}^2\bigg)\bigg(r_{0}^2-3r^2\bigg)\bigg)
+\bigg(69-16F^2\bigg)\bigg(\frac{16r^4F^6}{\bigg(r^2
+r_{0}^2\bigg)^{14}}\bigg)+\frac{r^2F}{\bigg(r^2+r_{0}^2\bigg)^{5}}\bigg(64F^2-76\bigg)\nonumber\\&&
+\frac{r^2}{F^2}\bigg(\frac{84}{F^2}+64\bigg)\bigg]+\bigg(\frac{r^2+r_{0}^2}{F}\bigg)^2+6\bigg(\frac{r^2+r_{0}^2}{F}\bigg)^7
+\frac{F^3}{\bigg(r^2+r_{0}^2\bigg)^7}\bigg(r^2\bigg(16F^2
-63\bigg)\nonumber\\&&
+\frac{9F}{2}\bigg(2F\bigg(5r^2+r_{0}^2\bigg)-2z_{2}\bigg(r^2+r_{0}^2\bigg)\bigg(r_{0}^2-3r^2\bigg)\bigg)\bigg).
\end{eqnarray}
The mass function (\ref{c5}), in this case reduces to,
\begin{eqnarray}\label{e4}
&&m\bigg(r,t\bigg)=\frac{3}{2}\bigg[\frac{F^2}{\bigg(r^2+r_{0}^2\bigg)^2}\bigg(1-\bigg(\frac{F'}{r^2+r0^2}
-\frac{2rF}{\bigg(r^2+r0^2\bigg)^2}\bigg)^2\bigg(\frac{F}{r^2+r0^2}\bigg)^5
+\frac{\bigg(r^2+r0^2\bigg)^5}{F^5}\bigg)\nonumber\\&&
+2\alpha\bigg(1-\bigg(\frac{F'}{r^2+r0^2}-\frac{2rF}{\bigg(r^2
+r0^2\bigg)^2}\bigg)^2\bigg(\frac{F}{r^2+r0^2}\bigg)^5+\frac{\bigg(r^2+r0^2\bigg)^5}{F^5}\bigg)^2\bigg].
\end{eqnarray}
With the help of Eq.(\ref{e2}) and Eq.(\ref{e3}), the Eq.(\ref{b3}) take the form
\begin{eqnarray}\label{e5}
&&\Delta a=1-\frac{E_{1}+E_{2}}{E_{3}+E_{4}}
\end{eqnarray}
where\\
$E_{1}=\alpha\bigg[60\bigg(\frac{r^2+r_{0}^2}{F}\bigg)^9+110\frac{\bigg(r^2+r_{0}^2\bigg)^8}{F^{14}}
+\bigg(\frac{10F^2}{\bigg(r^2+r_{0}^2\bigg)^5}-\frac{10}{F^3}-\frac{60r^2F^7}{\bigg(r^2
+r_{0}^2\bigg)^{14}}\bigg)\bigg(2F\bigg(5r^2+r_{0}^2\bigg)
-2z_{2}\bigg(r^2+r_{0}^2\bigg)\bigg(r_{0}^2-3r^2\bigg)\bigg)+\bigg(69-16F^2\bigg)\bigg(\frac{16r^4F^6}{\bigg(r^2
+r_{0}^2\bigg)^{14}}\bigg)+\frac{r^2F}{\bigg(r^2+r_{0}^2)^{5}}\bigg(64F^2-76\bigg)+\frac{r^2}{F^2}\bigg(\frac{84}{F^2}
+64\bigg)\bigg]$,\\
$E_{2}=\bigg(\frac{r^2+r_{0}^2}{F}\bigg)^2+6\bigg(\frac{r^2+r_{0}^2}{F}\bigg)^7
+\frac{F^3}{\bigg(r^2+r_{0}^2\bigg)^7}\bigg(r^2\bigg(16F^2-63\bigg)
+\frac{9F}{2}\bigg(2F\bigg(5r^2+r_{0}^2\bigg)\\-2z_{2}\bigg(r^2+r_{0}^2\bigg)\bigg(r_{0}^2-3r^2\bigg)\bigg)\bigg)$,\\
$E_{3}=6\alpha\bigg[5\bigg(\frac{r^2+r_{0}^2}{F}\bigg)^9\bigg(1+\bigg(\frac{r^2+r_{0}^2}{F})^5\bigg)
-4r^2\bigg(\frac{25}{F^4}-\frac{5F}{\bigg(r^2+r_{0}^2\bigg)^5}\bigg(1-\frac{4r^2F^5}{\bigg(r^2+r_{0}^2\bigg)^9}\bigg)\bigg]$\\
and\\
$E_{4}=\frac{3}{4}\bigg[-2\frac{\ddot{z_{2}}}{\dot{z_{2}}}\bigg(\frac{r^2+r_{0}^2}{F}\bigg)^{\frac{19}{2}}\bigg(5
+2\bigg(\frac{F}{r^2+r_{0}^2}\bigg)^{\frac{7}{2}}\bigg)+\frac{56r^2F^3}{\bigg(r^2+r_{0}^2\bigg)^7}
-25\bigg(\frac{r^2+r_{0}^2}{F}\bigg)^\frac{21}{2}
-4\bigg(\frac{r^2+r_{0}^2}{F}\bigg)^7-4\bigg(\frac{r^2+r_{0}^2}{F}\bigg)^2\bigg]$.\\

In this case, we take $r_{0}=0.5$, $z_{2}=1+t^2$, and the various values of $\alpha$, and analyzed our results. For $\beta=-\frac{5}{2}$, we obtained $\Theta>0$, the \textbf{Fig(7)} shows that, for the variation of $\alpha$ the energy density increases for the time profile. The radial pressure is initially maximum and then decreasing along $t$, the tangential pressure is decreasing near the center along $t$ at different values of $\alpha$ as shown in \textbf{Fig(8)} and \textbf{(9)}. The mass of the sphere is increasing along $t$ and the different values of $\alpha$ as shown in \textbf{Fig(10)}. In this case $p_{t}>p_{r}$, this implies that $\Delta a>0$. The anisotropy parameter increases in this case for the different values of $\alpha$ as shown in \textbf{Fig(11)}.shows that, firstly the anisotropy decreases and then increases with radial increase of star for various values of $\alpha$. The energy conditions for the expansion case are also satisfied as shown in \textbf{Fig(12)}.

\begin{figure}
\begin{center}
\includegraphics[width=70mm]{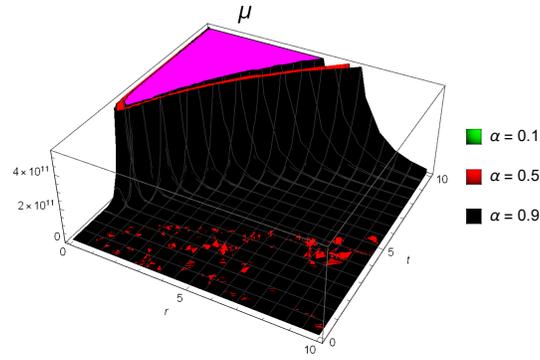}
\caption{Density behavior along $r$ and $t$ for $r_{0}=0.5$ and the different values of $\alpha$.}
\end{center}
\end{figure}
\begin{figure}
\begin{center}
\includegraphics[width=70mm]{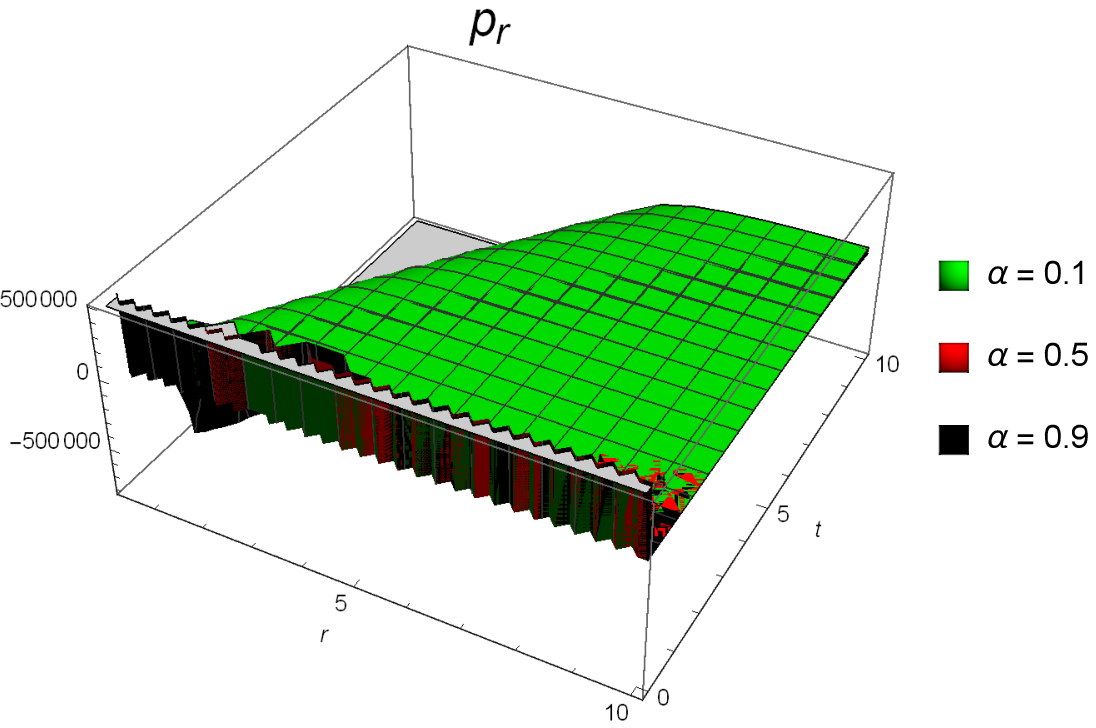}
\caption{Radial pressure behavior along $r$ and $t$ for $r_{0}=0.5$ and the different values of $\alpha$.}
\end{center}
\end{figure}
\begin{figure}
\begin{center}
\includegraphics[width=70mm]{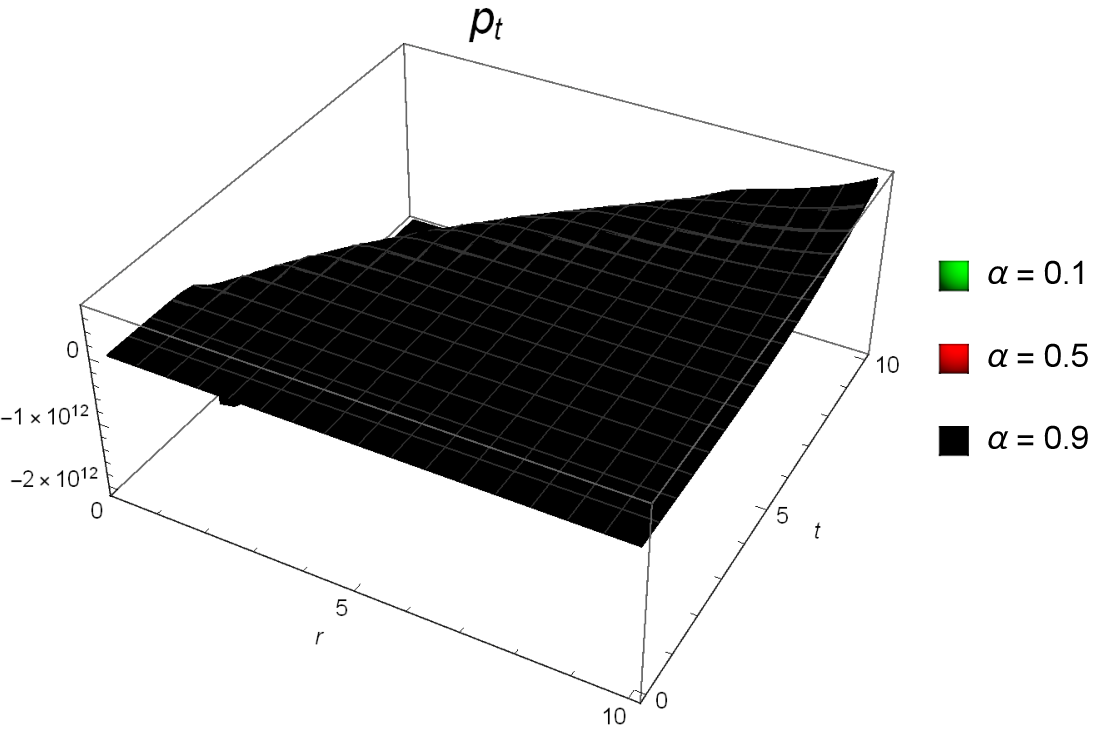}
\caption{Tangential pressure behavior along $r$ and $t$ for $r_{0}=0.5$ and the different values of $\alpha$.}
\end{center}
\end{figure}
\begin{figure}
\begin{center}
\includegraphics[width=70mm]{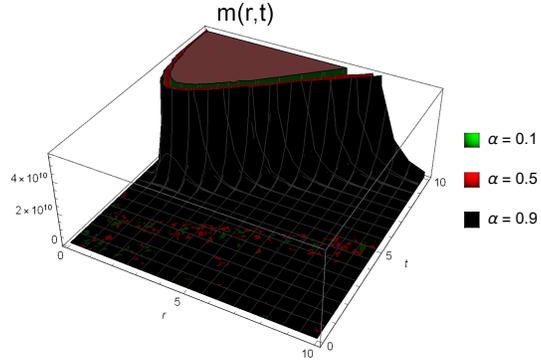}
\caption{Mass behavior along $r$ and $t$ for $r_{0}=0.5$ and the different values of $\alpha$.}
\end{center}
\end{figure}
\begin{figure}
\begin{center}
\includegraphics[width=70mm]{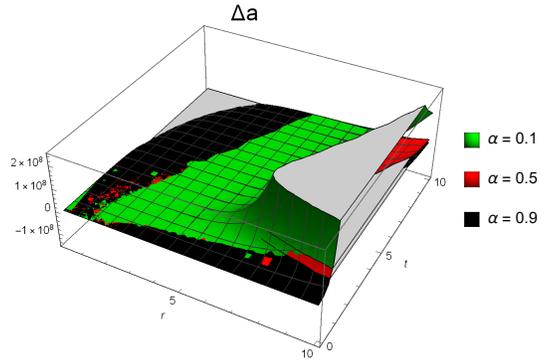}
\caption{Anisotropy parameter behavior for $r_{0}=0.5$ and the different values of $\alpha$.}
\end{center}
\end{figure}
\begin{figure}
\begin{center}
\includegraphics[width=60mm]{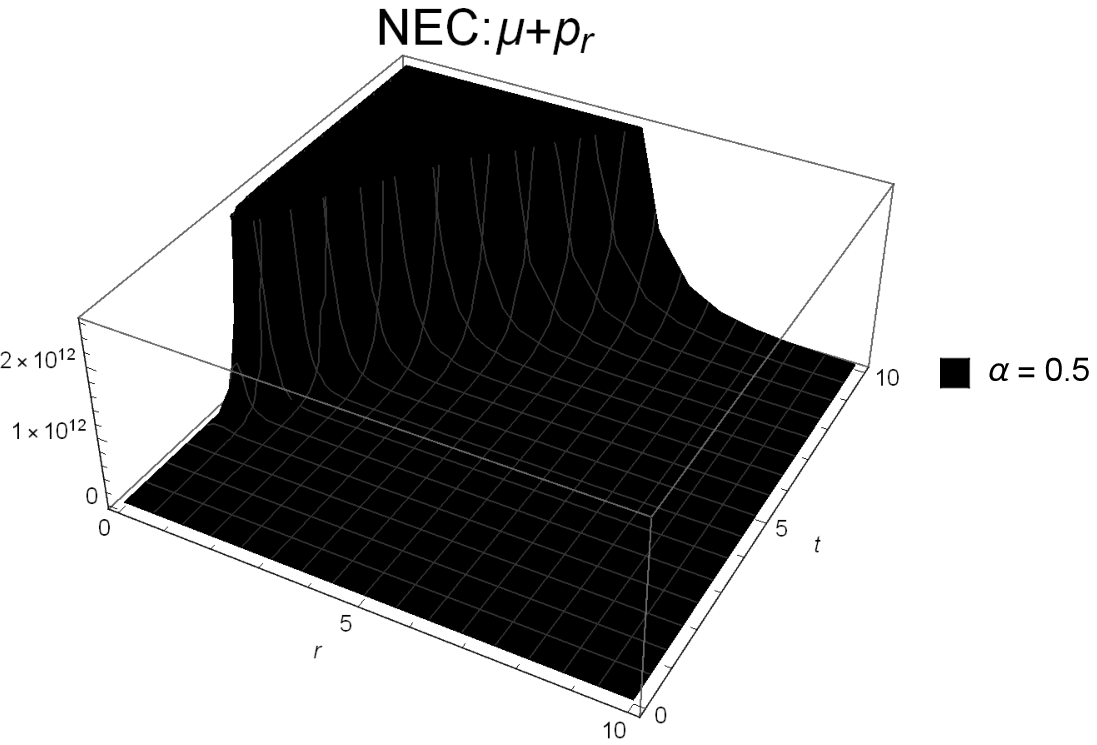}
\includegraphics[width=60mm]{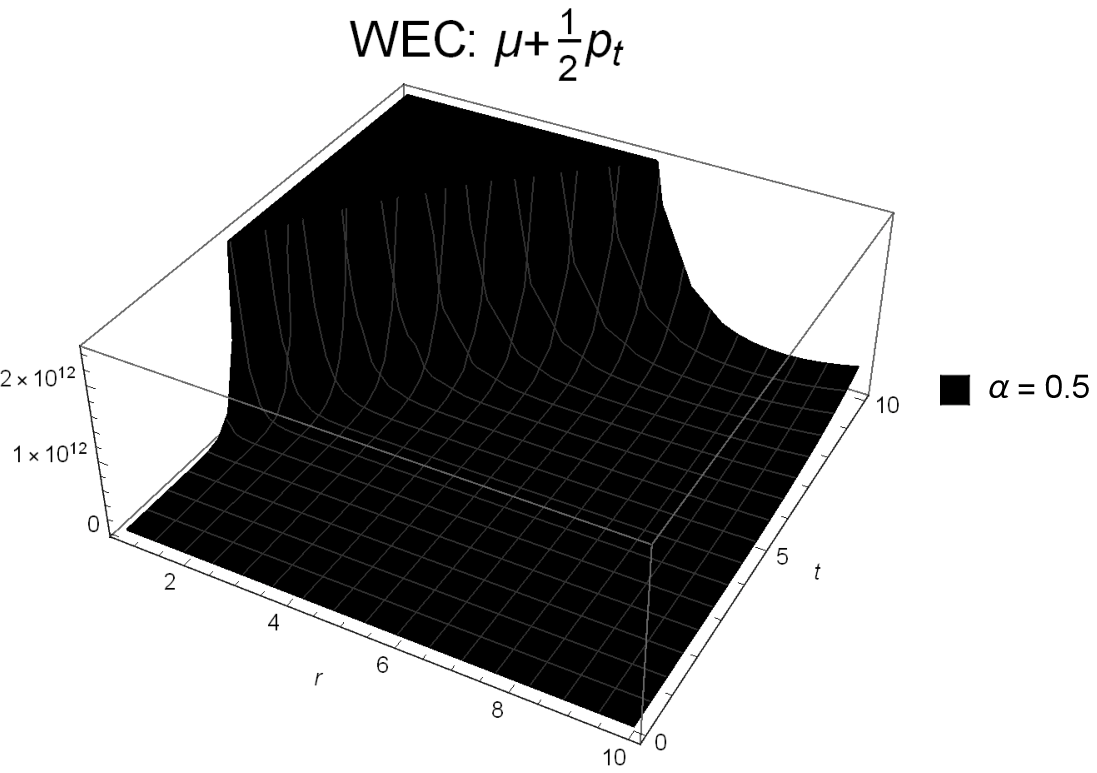}
\includegraphics[width=60mm]{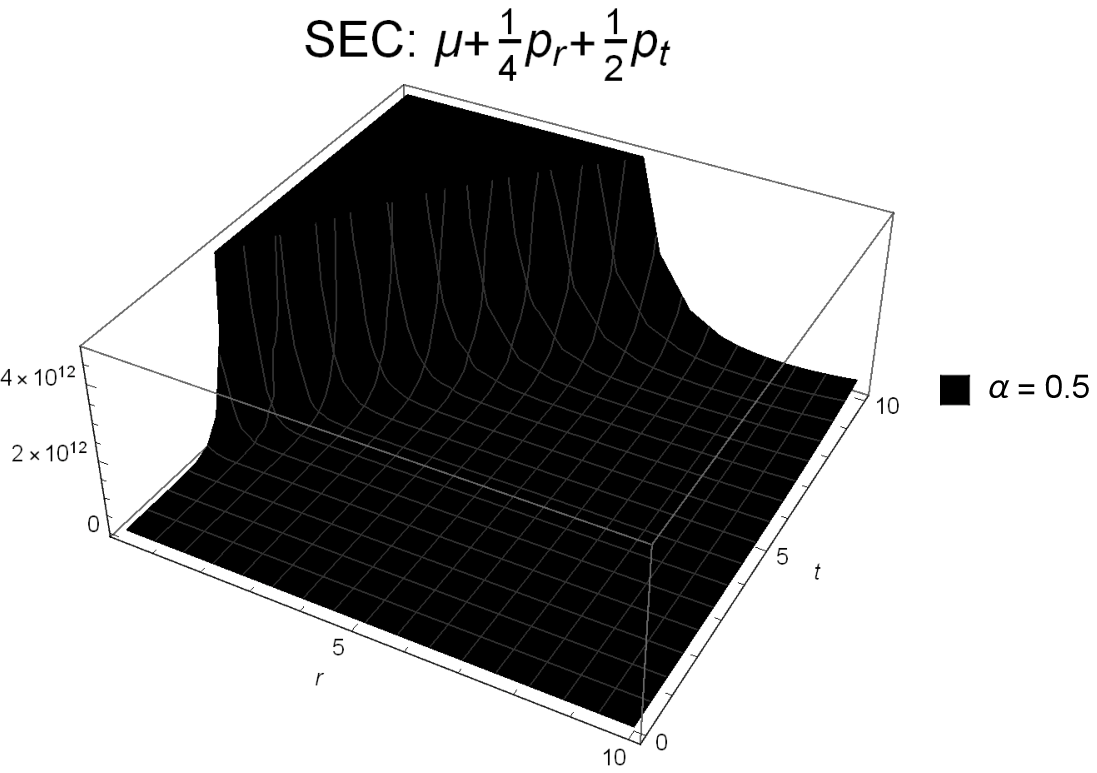}
\caption{Plot of energy conditions along $r$ and $t$ for $r_{0}=0.5$ at.}
\end{center}
\end{figure}
\section{Summary}
There has been a great interest to study the relativistic anisotropic systems due to the existence of such systems in the astronomical objects. The exact solutions of anisotropic sphere are helpful to determine the anisotropy of the universe during any era. Barrow and Maartens \cite{pp} studied the effects of anisotropy on the late time expansion of inhomogeneous universe. Mahmood et al. \cite{pp1} modeled the exact solution for the gravitational collapse and expansion of charged anisotropic cylindrical source.\\
The EGB gravity theory is the low energy limits of super-symmetric string theory of gravity \cite{e1}. The gravitational collapse is highly dissipative process in which a lot amount of energy is released \cite{33}. The dynamic of an anisotropic fluid collapse is observed in 5-Dimensional Einstein Gauss-Bonnet gravity \cite{g1}. Collin \cite{e2} modeled the in-homogeneous cosmological non-static expanding solutions. For the suitable values of $R(t,r)$ and $\beta$ Glass \cite{65} observed anisotropic collapsing and expanding of spherical object. We have modeled the field equations of anisotropic fluid in $5D$ EGB gravity.\\
 The aim of this paper is to study the generating solutions for anisotropic spherical symmetric fluid in $5D$ EGB theory of gravity. We have used auxiliary solution of one field equation to obtained the solutions for remaining field equations. The solution for expansion scalar $\Theta$ have been depending on the range of free constant $\beta$, for which $\Theta$ is positive or negative, leads to expansion and collapse of the fluid. We use the condition $\acute{R}=R^{2\beta}$ in Eq.(\ref{c5}), which leads to two trapped horizons at $R=\pm\sqrt{\frac{2}{3}m-2\alpha}$ , provided $m\geq3\alpha$. The curvature singularity is hidden at the common center of the inner $(R_{+})$ and outer $(R_{-})$ horizon. The matter components like density, radial pressure, tangential pressure, anisotropic parameter and mass functions have been determined in the $5D$ EGB theory of gravity.  The Eq.(\ref{b9}) implies that for $\beta=-3$, $\Theta=0$, for $\beta<-3$ the expansion scalar is negative and for $\beta>-3$ the expansion scalar is positive, which leads to bouncing, collapsing and expanding respectively. in other words, $\beta>0$, $\beta\in(-3, \infty)$ and $\beta<0$, $\beta\in(-\infty, -3)$. The solutions have been modeled by taking $\beta=-\frac{7}{2}$ and $\beta=-\frac{5}{2}$ for collapse and expansion of gravitating source respectively.\\
We assume the areal radius $R(r,t)=k(8r+z(t))^{\frac{1}{8}}$ out side the sphere for $k>1$. It is interesting that, for $k=1$, $R(r,t)=R_{trap}$ and is valid only for trapped surface, so we take $k>1$, for collapsing solutions outside the trapped surface. Further, $k<0$ should not be considered because this leads to the solutions corresponding to the inner surface of the trapped region, which is not the case of interest in the present discussion.\\
The dynamics of the spherical fluid is discussed in both cases. The density of the matter is decreasing/increasing in collapse/expansion with the arbitrary choice of constant/parameter, time profile and different values of GB term $\alpha$. The radial pressure, tangential pressure and mass have different behavior in both cases. The anisotropy is increasing in both cases, in other words, this pressure anisotropy in non-vanishing in both cases, leads to collapse and expansion of the fluid. The energy conditions are satisfied for collapse and expansion, this shows that our solutions are physically acceptable.

\section*{Acknowledgment}
One of us G. Abbas appreciates the financial support from HEC, Islamabad, Pakistan under NRPU project with grant number 20-4059/NRPU/R \& D/HEC/14/1217.
\section*{Conflict of Interest}
The authors declares that there is no conflict of interest regarding the publication of this manuscript.  
\section*{Data Statement}
The data used for this research is available with corresponding author and can be provided on request.


\vspace{0.25cm}
\begin{thebibliography}{40}
\bibitem{1} T. Kaluza, P.A.W.(1921)P. 966.
\bibitem{2} O. Klein, Z. Phys. \textbf{37}(1926)895; Nature \textbf{118},516(1926).
\bibitem{3} E Witten, Nucl. Phys. \textbf{B186},412(1981).
\bibitem{4} M. B. Green and J. H. Schward, Phys. Lett. \textbf{149B},117(1984).
\bibitem{5} D. J. Gross, J. Harvey, E. Martinec and R. Rohm, Phys. Rev. Lett. \textbf{6},502(1985).
\bibitem{7} Bruno ZUMINO, Phys. Rev. Lett. \textbf{137, No. 1},(1986)109-114.
\bibitem{8} D. G. Boulware and S. Deser, Phy. Rev. Lett. \textbf{55},2656(1985).
\bibitem{9} B. Zwiebach, Berkeley Preprint UCB.PTH-85/10.
\bibitem{10} Pedro. et al, arXiv: 1701.00079v2 [gr-qc].
\bibitem{11} G. Abbas and S. Sawar, Astro. Phys. Space Sci. \textbf{357}, 23(2015).
\bibitem{12} G. Abbas and M. Zubair, Mod. Phys. Lett. A \textbf{30},1550038(2015).
\bibitem{13} D. T. Gross and J.H. Sloan, Nucl. Phys. \textbf{B29}, 41(2006); M. C. Bento and O. Bertolami, Phys.Lett. \textbf{B368},198(1996).
\bibitem{14} N. Dadhich, A. Molina, A. Khugaev, Phys. Rev. \textbf{D81},104026(2010).
\bibitem{15} I. Antoniadis, J. Rizos and K. Tamvakis, Nucl. Phys. \textbf{B415}, 497(1994).
\bibitem{16} P. Kanti, J. Rizos and K. Tamvakis, Phys. Rev.  \textbf{D59}, 083512(1999).
\bibitem{17} P. Kanti, N. E. Mavromatos, J. Rizos, K. Tamvakis and E. Winstanley, Phys. Rev. \textbf{D54}, 5049(1996); \textbf{D57},6255(1998).
\bibitem{18} T. Torii, H. Yajima and K.I. Maeda, Phy. Rev. \textbf{D55}, 739(1997).
\bibitem{19} F. Tangherlini, Nuovo Cimento \textbf{27}, 636(1963).
\bibitem{20} R. C Myers and M. J. Perry, Annals of Phys. \textbf{172}, 304(1986)
\bibitem{21} J. T. Wheeler, Nucl. Phys. \textbf{B268}, 737(1986).
\bibitem{22} T. Torii and H. Maeda, Phys. Rev. \textbf{D71}, 124002(2005).
\bibitem{23} R. C. Myers and J. Z. Simon, Phys. Rev. \textbf{D38}, 2434(1988).
\bibitem{24} R. G. Cai, Phys. Rev. \textbf{D65}, 084014(2002); R. G. Cai and Q. Guo, Phys. Rev. \textbf{D69}, 104025(2004).
\bibitem{25} T. Kobayashi, Gen. Rev. Grav. \textbf{37}, 1869(2005).
\bibitem{26} H. Maeda, Class. Quantum Grav. \textbf{23}, 2155(2006); Phys. Rev. \textbf{D73}, 104004(2006) .
\bibitem{27} A. E. Dominguez and E. Gallo, Phys. Rev. \textbf{D73}, 064018(2006).
\bibitem{28} S. G. Ghosh and D. W. Deshkar, Phys. Rev. \textbf{D77}, 047504(2008).
\bibitem{29} S. W. Hawking and G. F. R. Ellis, The Large Scale Structure of Space time (Cambridge University Press, Cambridge, 1979).
\bibitem{30} L. Herrera, N. O. Santos, G. Le Denmat,MNRAS \textbf{237}, 257(1989).
\bibitem{31} C. W. Misner and D. Sharp: Phys. Rev. \textbf{136B}, 571(1964).
\bibitem{32} C. W. Misner and D. Sharp: Phys. Rev. \textbf{137B}, 1360(1965).
\bibitem{33} L. Herrera and N. O. Santos: Phys. Repot \textbf{286}, 53(1997).
\bibitem{34} L. Herrera, A. Di Prisco and J. R Hernan Dezand, N. O. Santo: Phys. Lett. A \textbf{237}, 113(1998).
\bibitem{35} L. Herrera, N. O. Santos: Phys. Rev. \textbf{D70}, 084004(2004).
\bibitem{36} L. Herrera, A. Di Prisco and J. Ospino: Gen. Relativ. Gravit. \textbf{44}, 2645(2012).
\bibitem{37} L. Herrera: Int. J. Mod. Phys. \textbf{D15}, 2197(2006).
\bibitem{38} L. Herrera, N. O. Santos and A. Wang: Phys. Rev. \textbf{D78}, 084024(2008).
\bibitem{39} G. Abbas, sci. China. Phys. Mechanic Astro \textbf{57}, 604(2014).
\bibitem{40} S. M. Shah and G. Abbas, Eur. Phys. J. C \textbf{77}, 251(2017).
\bibitem{41} G. Abbas, AstroPhys. Space Sci. \textbf{350}, 307(2014).
\bibitem{42} G. Abbas, Adv. High Energy Phys. \textbf{2014}, 306256(2014).
\bibitem{43} G. Abbas, Astrophys. Space Sci. \textbf{352}, 955(2014).
\bibitem{44} G. Abbas and U. Sabiullah, AstroPhys. Space Sci. \textbf{352}, 769(2014).
\bibitem{b}  M. Zubair, Hina Azmat, I. Noureen: Eur. Phys. J. \textbf{C77}, 169(2017).
\bibitem{Ch1} K.Zhou, Z-Y., Yang, De-Cheng Zou and R-H., Yue, Int. J. Mod. Phys. \textbf{D22}, 2317(2011).
\bibitem{Ch2} K. Zhou, Z-Y., Yang, De-Cheng Zou and R-H., Yue, Mod. Phys. Lett. \textbf{A26}, 2135(2011).
\bibitem{Ch3} R-H., Yue, De-Cheng Zou, T-Y., Yu, and Z-Y., Yang, Chin. Phys. \textbf{B20}, 050401(2011).
\bibitem{Ch33}  K.Zhou, Z-Y., Yang, De-Cheng Zou and R-H., Yue,  Chin. Phys. \textbf{B21}, 020401(2012)
\bibitem{Ch4} De-Cheng Zou, Z-Y., Yang, R-H., Yue, and T-Y., Yu, Chin. Phy.\textbf{B20} ,100403(2011).
\bibitem{Ch5} R-H., Yue, De-Cheng, Zou, T-Y., Yu, P. Li and Z-Y., Yang, Gen. Relativ. Gravit.\textbf{43}, 2103(2011).
\bibitem{Ch6} De-Cheng, Zou, Z-Y., Yang, and R-H., Yue, Chin. Phys. Lett.\textbf{28}, 020402(2011).
\bibitem{Ch7} De-Cheng Zou, Zhan-Ying Yang, Rui-Hong Yue, P. Li, Mod. Phys. Lett. \textbf{A26}, 515(2011).
\bibitem{Ch8} De-Cheng Zou, R-H., Yue, and Z-Y., Yang, Commun. Theor. Phys.\textbf{55}, 499(2011).
\bibitem{c}  S. Chakraborty: Relativ. Gravit. \textbf{45}, 2039(2013).
\bibitem{45} G. Abbas and Riaz Ahmad.Eur. Phys. .J . \textbf{C441}, 77(2017).
\bibitem{45a} M. Sharif and A. Siddiqa, Int. J. Mod. phys. \textbf{D18}, 1950005(2018).
\bibitem{46} J. R. Oppenheimer and H. Snyder, Phys. Rev. \textbf{56}, 455(1939).
\bibitem{50} D. Markovic and S. L. Shapiro, Phy. Rev. \textbf{D61}, 084029(2000).
\bibitem{51} K. Lake, Phys. Rev. \textbf{D62}, 027301(2000).
\bibitem{52} M. Sharif and G. Abbas, Astrophys. Space Sci. \textbf{327}, 285-291(2010).
\bibitem{53} M. Sharif and Zahid Ahamd, Mod. Phys. Lett. \textbf{A22}, 1493(2007).
\bibitem{54} M. Sharif and Zahid Ahamd, Mod. Phys. Lett. \textbf{A22}, 2947(2007).
\bibitem{55} M. Sharif and Zahid Ahamd, J. Korean Phys. Society. \textbf{52}, 980(2008).
\bibitem{56} M. Sharif and Zahid Ahamd, Acta Phys. Polonica. \textbf{B39}, 1337(2008).
\bibitem{57} M. Sharif and G. Abbas, J. Korean Physical Society \textbf{56}, 529(2010).
\bibitem{g1} G. Abbas and M. Zubair, Arxiv: 1504.07937v1 (2015).
\bibitem{g2} D. Ida and K. I. Nakao, Phys. Rev. \textbf{D66}, 064026(2002).
\bibitem{59} S. Jhingan, Sushant G. Ghosh, Phys. Rev. \textbf{D81}, 024010(2010).
\bibitem{60} Sunil D. Maharaj, Brian Chilambwe and Sudan Hansraj, Phys. Rev. D91, 084049(2015).
\bibitem{61} G. Abbas and M. Tahir, Eur. Phys. J. C, \textbf{537}, 77(2017).
\bibitem{62} A. Banerjee, A. Sil and S. Chatterjee, Gen. Relativ. Gravit. \textbf{26}, 999(1994).
\bibitem{63} S. G. Ghosh and A. Banerjee, Int. J. Mod. Phys. \textbf{D12}, 639(2003).
\bibitem{64} S. G. Ghosh, D. W. Deshkar and N. N. Saste, Int. J. mod. Phys. \textbf{D16}, 53(2007).
\bibitem{65} E. N. Glass. Gen. Relativ. Gravit. \textbf{45}, 266(2013).
\bibitem{pp} J. D. Barrow and R. Maartens: Phys. Rev. \textbf{D59}, 043502(1998).
\bibitem{pp1} T. Mahmood, S. M. Shah and G. Abbas: AstroPhys. space. sci. \textbf{6}, 2234(2015).
\bibitem{e1} J. J. Schwarz, Nucl. Phys. \textbf{B226}, 269(1983).
\bibitem{e2} C. B. Collins: J. Math. Phys. \textbf{18}, 2116(1977).

\end{thebibliography}
\end{document}